# Advances in imaging THGEM-based detectors

M. Cortesi[a], R. Chechik[a,*], A. Breskin[a], G. Guedes[a],

V. Dangendorf[b], D. Vartzky[c], and D. Bar[c]

[a] *Dept. of Particle Physics, Weizmann institute of science, 76100 Rehovot, Israel*

[b] *Physikalisch-Technische Bundesanstalt (PTB), Braunschweig, Germany*

[c] *SOREQ NRC, Yavne, Israel*

**Abstract**

We report on recent measurements with Thick GEM-like (THGEM) – based imaging detectors. The THGEM is a robust gaseous electron multiplier similar to GEM but with larger dimensions. It has high electron multiplication, of $10^5$ and $10^7$ in single- and double-THGEM structure, respectively, fast signals and ~10MHz/mm$^2$ rate-capability. It can be produced in any shape and over large area. In view of many possible applications of THGEM-based imaging detectors, in particle physics and beyond, we have recently studied the localization properties of a 2D 10x10 cm$^2$ detector; the results are shortly presented.

## 1. Introduction

The thick GEM (THGEM) [1] is an "expanded" GEM, economically produced in the PCB industry by simple drilling and etching in G-10 or other insulating materials (fig. 1). Similar to GEM, its operation is based on electron gas avalanche multiplication in sub-mm holes, resulting in very high gain and fast signals. Due to its large hole size, the THGEM is particularly efficient in transporting the electrons into and from the holes, leading to efficient single-electron detection and effective cascaded operation.

The THGEM provides true pixilated radiation localization, ns signals, high gain and high rate capability. For a comprehensive summary of the THGEM properties, the reader is referred to [2, 3]. In this article we present a summary of our recent study on THGEM-based imaging, carried out with a 10x10 cm$^2$ double-THGEM detector.

## 2. The 10x10 cm$^2$ imaging detector

The detector was made of two THGEMs (10x10 cm$^2$) in cascade, coupled to a resistive anode made of 2 MΩ/□ graphite on G10 substrate. Behind the resistive anode the broadened induced signals were captured on a double-sided X-Y readout electrode, structured with pad strings with a pitch of 2 mm and equipped with discrete delay-line circuits for each coordinate. The detector was operated in atmospheric Ar/CH$_4$ (95:5) and its performance was studied with 6-8 keV X-rays. The detector yielded ~20% FWHM local energy resolution and gain uniformity of ±10% FWHM over the whole surface.

The imaging resolutions were studied with a steel mask, having slits series of frequencies 0.05 to 1 line-pairs per mm. The contrast transfer function, CTF, was calculated from the recorded image (fig. 3). The spatial resolution is better then 0.4 mm FWHM, in spite of the 1mm hole pitch; it has good linearity and good homogeneity. Furthermore, the image has very low digital noise, of the order of ~3 % standard deviation across the sensitive area.

THGEM-based detectors could become an attractive robust and economic solution for numerous applications requiring large-area detectors with sub-mm localization resolution. Examples are: elements for large TPC readout, sampling elements in Calorimetry (e.g. for ILC), UV-photon imaging in RICH [4], LXe scintillations recording, moderate-resolution tracking (LHC2), x-ray and neutron imaging, etc. Of particular interest is the detection of UV photons in RICH systems, presently performed with GEM, with the possibility to deposit the photocathode on the THGEM's top face and operate it in a mode insensitive to the ionizing hadronic background [4].

The work was supported by the Benoziyo Center for High Energy Research and the Israel Science Foundation. A.B. is the W.P. Reuther Professor of Research in the peaceful use of Atomic Energy.

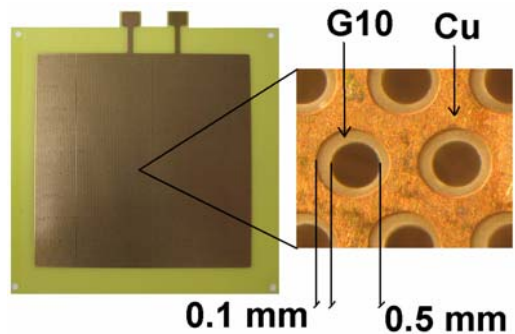

Figure 1. 10x10 cm$^2$ THGEM, 0.4 mm thick, with a 0.5mm hole diameter and 1 mm pitch. The enlarged picture shows the details of the THGEM's etched rim.

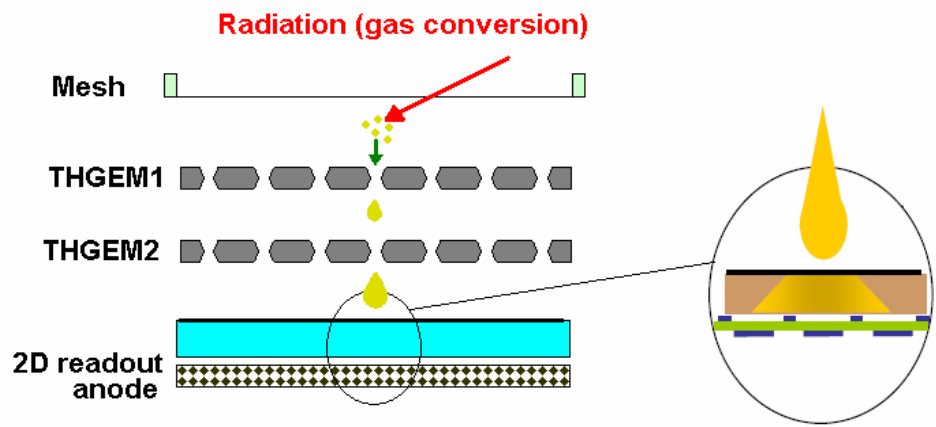

Figure 2. A scheme of the imaging detector comprising two THGEMs, a resistive anode, and a double-sided pickup electrode structured with 2D pad-strings connected to delay-lines.

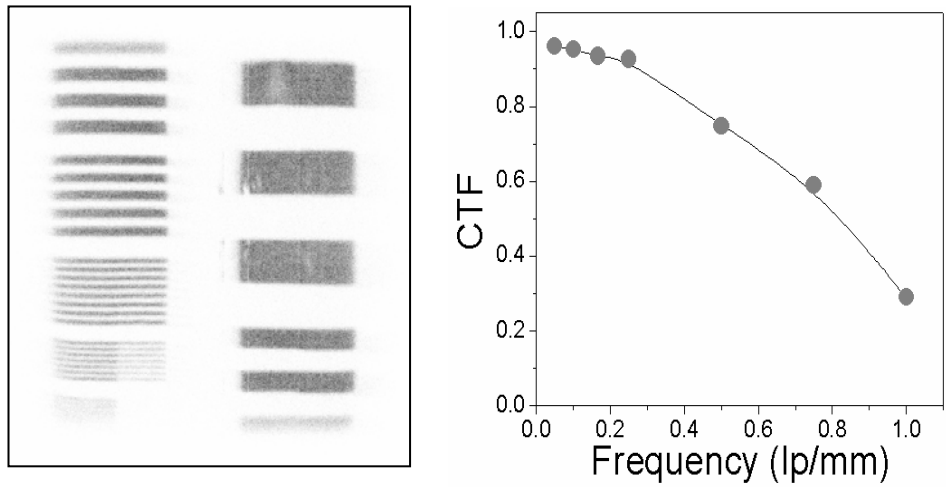

Figure3. Contrast Transfer Function (CTF) obtained from a steel mask with slits' frequency of 0.05 to 1 line pairs per mm.